\documentclass[pra,10pt,twocolumn,superscriptaddress,nofootinbib, amsmath,amssymb,
floatfix]{revtex4-2}

\usepackage{bm,braket}
\usepackage{bbold,dsfont}
\usepackage{units}
\usepackage{amssymb}
\usepackage{amsthm}
\usepackage{mathtools}
\usepackage{mathrsfs}

\usepackage{graphicx}
\usepackage{comment}
\usepackage{subcaption}

\usepackage{natbib,hyperref}
\usepackage[normalem]{ulem}
\usepackage{xfrac} 
\usepackage{dsfont}
\usepackage{lipsum}

\newtheorem{theorem}{Theorem}

\newtheorem{lemma}[theorem]{Lemma}

\def\>{\rangle} 
\def\<{\langle}

\usepackage{xcolor}

\DeclareMathOperator{\tr}{tr}

\begin{document} 

\title{Approximate reconstructability of quantum states\\ and noisy quantum secret sharing schemes}

\author{Yingkai Ouyang}
\email{y.ouyang@sheffield.ac.uk}
\affiliation{Department of Physics \& Astronomy, University of Sheffield, Sheffield, S3 7RH, United Kingdom}
\affiliation{Centre for Quantum Technologies, National University of Singapore, Singapore}

\author{Kaumudibikash Goswami}
\affiliation{School of Mathematics and Physics, University of Queensland, Brisbane, Queensland 4072, Australia}
\affiliation{Raman Research Institute, 
Sadashivanagar, Bengaluru, Karnataka 560080, India}

\author{Jacquiline Romero}
\affiliation{School of Mathematics and Physics, University of Queensland, Brisbane, Queensland 4072, Australia}

\author{Barry C. Sanders}
\email{sandersb@ucalgary.ca}
\affiliation{Institute for Quantum Science and Technology, University of Calgary, Alberta T2N 1N4, Canada}
\affiliation{Raman Research Institute, 
Sadashivanagar, Bengaluru, Karnataka 560080, India}
\author{Min-Hsiu Hsieh}
\email{min-hsiu.hsieh@foxconn.com}
\affiliation{Hon Hai (Foxconn) Research Institute, Taipei, Taiwan}

\author{Marco Tomamichel}
\affiliation{Centre for Quantum Technologies, National University of Singapore, Singapore}
\affiliation{Department of Electrical and Computer Engineering, National University of Singapore, Singapore 117583, Singapore}

\begin{abstract} 
We introduce and analyse approximate quantum secret sharing in a formal cryptographic setting, wherein a dealer encodes and distributes a quantum secret to players such that authorized structures (sets of subsets of players) can approximately reconstruct the quantum secret and omnipotent adversarial agents controlling non-authorized subsets of players are approximately denied the quantum secret. In particular, viewing the map encoding the quantum secret to shares for players in an authorized structure as a quantum channel, we show that approximate reconstructability of the quantum secret by these players is possible if and only if the information leakage, given in terms of a certain entanglement-assisted capacity of the complementary quantum channel to the players outside the structure and the environment, is small. 
\end{abstract}

\maketitle

\section{Introduction}

Quantum resources enable cryptographic tasks beyond what is classically possible. For instance, quantum key distribution~\cite{BB84,Eke91} provides an information-theoretic means for generating shared classical keys. Secret sharing (SS) is another fundamental cryptographic primitive, wherein a dealer~D distributes a secret
as shares to a set of players~$\wp$ such that any group in the authorised structure~$\Gamma\subseteq2^\wp$ (sets of authorised subsets of the players) reconstructs the secret by combining shares and decoding,
whereas groups in the complementary adversarial structure~$\bar\Gamma=2^\wp \setminus \Gamma$
cannot obtain any information about the secret.
SS has been quantised in two ways:
quantum-safe classical SS~\cite{hillery_quantum_1999}
and the version we employ here---quantum-secret sharing (QSS)~\cite{CGL99}
as a special case of quantum error correction~\cite{KnL97}---which can be partially unified
via quantum graph states for qubits~\cite{markham2008}
and subsequently for qudits~\cite{KFMS10}. 
Quantum secret sharing has applications in quantum Byzantine agreements~\cite{quantum-byzantine-prl}
and distributed quantum computation~\cite{Ouyang_QSS_2017}, amongst others.

\begin{figure}
\centering
\includegraphics[width=0.45\textwidth]{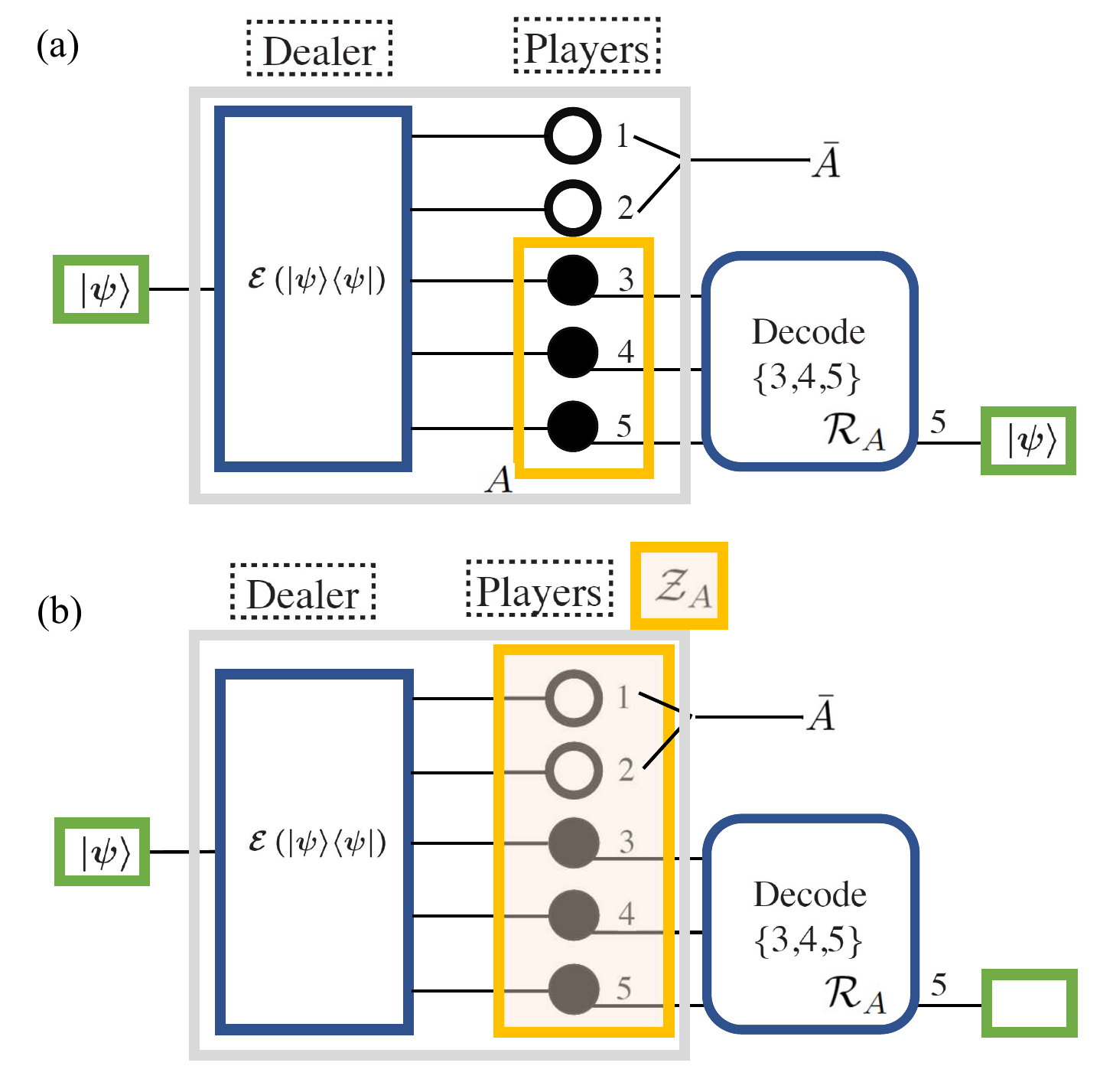}
\caption{
\textbf{(a)~Ideal threshold QSS scheme.}
The dealer encodes the secret with channel $\mathcal E$, and distributes the shares to players 1,2,3,4 and 5. Players in the set $A = \{3,4,5\}$ collaborate in the decoding using the map $\mathcal R_A$ and reconstruct the secret. 
We label the players outside $A$ as $\bar A = \{1,2\}$.
\textbf{(b)~Adversarial attack on a threshold QSS scheme.} The adversary colludes with players 1 and 2. They apply the map $\mathcal Z_A$ on the players' qudits, potentially adding noise to the systems of any player. Depending on the attack, the legitimate players can still approximately recover the secret $|\psi\>$. 
\label{fig:QSS}
}
\end{figure}

Ideal $(t,n)$-threshold QSS features perfect reconstructability
and perfect secrecy as elucidated in Fig.~\ref{fig:QSS}(a);
i.e., any~$t$ out of $n$ players can reconstruct the secret perfectly,
and perfect secrecy means that fewer than~$t$ players do not gain any information about the secret.
From this foundation,
generalised QSS can be constructed from threshold QSS
by evenly or unevenly distributing shares to players~\cite{CGL99,Got00,imai_quantum_2003}.
In $(t,n)$-QSS~\cite{HBB99,CGL99,Got00}, a dealer
D employs an encoding map~$\mathcal{E}$
to encode a quantum secret
$\varrho\in\mathcal{D}
(\mathscr{H})$
(trace-class positive density operator)
into~$n$ $q$-dimensional qudits,
i.e., onto Hilbert space
$\mathscr{H}_q^{\otimes n}$
($n$-fold tensor product of $q$-dimensional Hilbert spaces).
Each share of one qudit is sent to one of~$n$ players,
such that~$\Gamma$ comprises all groups of at least~$t$ players and~$\bar\Gamma$ is the complement,
namely, all groups of fewer than~$t$ players.

Here, we construct a theory of approximate secrecy and reconstructability by introducing an adversary model
as shown in Fig.~\ref{fig:QSS}(b). In our model, the adversary structure comprises omnipotent adversaries
who are denied control over~$\Gamma$
but can collaborate with players in~$\bar\Gamma$.
Imperfect SS has been considered,
but strong assumptions on the adversary's capability are required~\cite{nikova2006threshold}.
In contrast, the dichotomy between reconstructability and secrecy is quite general and is inherently quantum due to the no-cloning principle~\cite{Par70,WZ82}, devoid of any classical analogue:
classically, the ability to copy a secret allows an authorised set to reconstruct the secret exactly but cannot provide a guarantee that an adversary who could have intercepted the communication cannot do the same.
Approximate QSS relaxes the requirements of perfect reconstructability for $\Gamma$ and perfect secrecy for $\bar \Gamma$.
Approximate quantum secret sharing schemes derived from quantum Reed-Solomon codes were investigated in~\cite{crepeau2005approximate}, but this leaves open the question of how more general approximate quantum secret sharing schemes perform. The dichotomy between approximate recoverability and approximate secrecy has also been investigated~\cite{imai2005information, Ogawa_yamamoto_secret,Spekkens_complementarity,hayden_approximate_2020}, 
but it remains unclear how these quantities relate to the maximum rate at which the secret is transmitted to the adversary.

\section{Main result}
Consider now a $(t,n)$-threshold QSS scheme where a $q$-dimensional secret is shared with players holding qudits ($d$-dimensional quantum systems). In our model, given any $A \in \Gamma$,
the adversary can attack all qudits after the dealer applies the encoding map $\mathcal E$ and prior to reconstruction.
The effect of the adversary's action amounts to applying an effective channel 
$\mathcal Z_{A}$.
Thus, the quantum channel mapping the quantum secret to the quantum state on $A$ just before reconstruction is 
\begin{align}
\mathcal{N}_A
=\operatorname{tr}_{\bar A} \circ \mathcal Z_{A}  \circ \mathcal E, \label{def:chanNA}
\end{align}
with $\operatorname{tr}_{\bar A}$ denoting the partial-trace that removes the players in $\bar A = \{1,\dots, n\} \setminus A$.
The $|A|$ authorised players then apply a recovery channel~$\mathcal{R}_A$ that maps the qudits labelled by $A$ to a single $q$-dimensional system.

We then define our $(t,n)$-threshold QSS scheme to be $\delta$-reconstructable if 
\begin{align}
\delta= 
\max_{A :  |A|\ge t} 
\min_
{\substack{\mathcal{R}_A}}
D_{\diamond}( \mathcal{R}_A \circ \mathcal{N}_A,  \mathcal{I} ) , \label{eq:BO-reconstructability}
\end{align}
where the reconstruction channels $\mathcal{R}_A$ is of the form above, $\mathcal{I}$ denotes the identity channel, and $D_{\diamond}$ denotes the diamond (or stabilised) norm distance between quantum channels (see below). Here the maximisation is over all authorised groups, but without loss of generality we can restrict to structures with $|A| = t$. The diamond norm distance between two channels $\mathcal{E}$ and $\mathcal{F}$ is defined as
\begin{align}
	D_{\diamond}(\mathcal{E},\mathcal{F}) = \!\!\!\!
	\max_{|\psi\rangle \in \mathcal{H} \otimes \mathcal{H}'}   \frac12 
	\big\| \mathcal{E} \otimes \mathcal{I} ( |\psi\rangle\!\langle\psi| ) - \mathcal{F} \otimes \mathcal{I} ( |\psi\rangle\!\langle\psi| )  \big\|_1 ,
\end{align}
where $\| \cdot \|_1$ is the Schatten $1$-norm and the optimisation goes over all auxiliary Hilbert spaces $\mathcal{H}'$. The use of a stabilised distance here is crucial as it ensures that arbitrary secrets can be restored, inclusive of their correlations with a quantum memory held by a third party.

Alternatively, we can replace $D_{\diamond}$ with a fidelity-based stabilised distance, namely
\begin{align}
	F_{\diamond}(\mathcal{E},\mathcal{F}) = 
	\min_{|\psi\rangle \in \mathcal{H} \otimes \mathcal{H}'}   
	F \big( \mathcal{E} \otimes \mathcal{I} ( |\psi\rangle\!\langle\psi| ),  \mathcal{F} \otimes \mathcal{I} ( |\psi\rangle\!\langle\psi| )  \big)
\end{align}
where $F$ is the Uhlmann fidelity, $F(\rho,\tau) = \|\sqrt{\rho}\sqrt{\tau} \|_1^2$. We say that the scheme is $\epsilon$-reconstructable in fidelity if
\begin{align}
	\epsilon = 1 - \min_{A : t \le |A|} \max_{\substack{\mathcal{R}_A}} F_{\diamond}( \mathcal{R}_A \circ \mathcal{N}_A,  \mathcal{I} ) .
\end{align}

We can relate the two notions of recoverability using Fuchs-van de Graaf inequalities, namely, for any quantum channel $\mathcal F$, we 
show in the Supplemental Material that
\begin{align}
    D_\diamond(\mathcal F , \mathcal{I} ) 
    \ge
    1 - F_{\diamond} ( \mathcal F,  \mathcal{I} )      
    \ge
    D_\diamond(\mathcal F , \mathcal{I} )^2.
    \label{eta-identity} 
\end{align}
From this we can immediately conclude that $\gamma$-recoverability in fidelity implies $\sqrt{\gamma}$-recoverability in diamond norm, and conversely $\delta$-recoverability in diamond norm implies also $\delta$-recoverability in fidelity.

Next, we establish the notion of approximate secrecy. For this, we need to introduce {\em complementary channels}~\cite{DeS03} for the channels $\mathcal{N}_A$, which intuitively model how much information the adversary retains after the attack. In particular, for a channel $\mathcal{N}_A$ we introduce its Stinespring isometry $\mathcal{U}$ and define $\hat{\mathcal{N}}_A = \tr_{A} \circ\ \mathcal{U}$, where $\tr_A$ is the partial trace removing the authorized set.
Here, if $\mathcal E$ has Kraus operators $E_i$, $\mathcal Z_A$ has Kraus operators $Z_{A,j}$, and the partial trace on $\bar A$ has Kraus operators $\<k_{\bar A}| \otimes I_{A}$,
where $I_A$ is the identity operator on the authorized set $A$, then $\mathcal N_A$ has Kraus operators $(\<k_{\bar A}| \otimes I_A) Z_{A,j} E_i$. 
Then define the operator $W = \sum_{i,j,k}|i,j,k\> \otimes ((\<k_{\bar A}| \otimes I_A)) Z_{A,j} E_i$. The map $\mathcal U$ is then defined as $\mathcal U(\rho) = W \rho W ^\dagger$.

With this, we say that a $(t,n)$-threshold QSS scheme has $\epsilon$-secrecy if 
\begin{align}
\epsilon=  1-
\min_{A :  |A|\ge t}  \max_{\sigma} 
F_{\diamond}( \hat{\mathcal N}_A , \mathcal V_{A,\sigma} ), 
 \label{eq:BO-secrecyleakage}
\end{align}
where $\mathcal V_{A,\sigma}$ is a preparation channel 
that prepares a {fixed} density matrix $\sigma$. 
Namely,
{$\mathcal V_{A,\sigma}$} traces out the qudits of the players in $A$ and prepares a quantum state described by the density matrix $\sigma$, where the output $\sigma$ does not contain any information about the input state, i.e., the input state is completely hidden. Hence, when $\hat{\mathcal N}_A=\mathcal V_{A,\sigma}$ for some $\sigma$, we have $\epsilon=0$: a condition for perfect secrecy. The other extreme case is when $\hat{\mathcal N}_A=\mathcal{I}$, i.e., all the information is leaking through $\hat{\mathcal N}_A$.
In this case it can be seen that $\epsilon=1$. 

Finally, we define the strength~$C$ of the adversarial model  for a $(t,n)$-threshold QSS scheme:
    \begin{align}
        C=\max_{A : |A| \ge t}C\bigl(\hat{\mathcal N}_A\bigr),
    \end{align}
where $C(\hat{\mathcal N}_A)$ is the entanglement-assisted classical capacity of $\hat{\mathcal N}_A$, which is defined for a channel $\mathcal{N}$ with input labeled by $X$ and output labeled by $Y$ as
\begin{align}
	C(\mathcal{N}) = \max_{|\psi\rangle \in \mathcal{H}_X \otimes \mathcal{H}_X} I(X: Y)_{\tau}
\end{align}
where $\tau = \mathcal{I} \otimes \mathcal{N}  ( |\psi\rangle\!\langle\psi| )$ and $I(X:Y)_\tau$ is the quantum mutual information evaluated for the state $\tau$. 
The mutual information itself
can be expressed in terms of the Umegaki relative entropy, denoted $D(\cdot\|\cdot)$, namely
\begin{align}
	I(X:Y)_\tau =  \min_{\rho_Y}D(\tau\| \rho_{X} \otimes \rho_Y) \,, 
\end{align} 
where $\rho_{X}$ and $\rho_Y$ are the marginals of $\tau$.
Using this, we can introduce a modified entanglement-assisted capacity, where $I(X:Y)_{\tau}$ is replaced by
\begin{align}
	\tilde{I}(X:Y)_\tau = -  \max_{\rho_Y}\log F(\tau, \rho_{X} \otimes \rho_Y) \,,
\end{align} 
which is a variant of the mutual information based on the sandwiched R\'enyi relative entropy of order $\nicefrac12$~\cite{muller2013quantum,wilde2014strong},
given by 
\begin{align}
 \widetilde D_\alpha ( \rho \| \sigma)
=\frac1{\alpha - 1}
\log \operatorname{tr}\left(  (
 \sigma^{\frac{1-\alpha}{2\alpha}})^\alpha 
 \rho
 \sigma^{\frac{1-\alpha}{2\alpha}})^\alpha 
 \right)^\alpha,
\end{align}
where $\rho$ and $\sigma$ are quantum states and $\alpha \neq 1$.
The corresponding generalized mutual information is 
${\tilde I_\alpha(X:Y)_\tau = \min_{\rho_Y} \widetilde D_{\alpha} ( \tau \| \rho_X \otimes \rho_Y)}$ \cite{gupta2015multiplicativity},
and $
\tilde I(X:Y)_\tau
=
\tilde I_{\sfrac 1 2}(X:Y)_\tau
$.
The quantity $C_\alpha(\mathcal N) = \max_{\tau} \tilde I_\alpha(X:Y)_\tau$ is a generalized entanglement assisted capacity because
$C(\mathcal N) = \lim_{\alpha \to 1}C_\alpha(\mathcal N)$. Next, we define the modified strength of the adversarial model as
$\tilde{C} = \max_{|A| \ge t}
C_{\sfrac 1 2}(\hat{\mathcal N}_A)$, which corresponds to setting $\alpha=\sfrac 1 2$. The value of $\alpha=\sfrac 1 2$ is chosen to express the generalized mutual information in terms of fidelity. Since $ \widetilde D_\alpha$ is monotone nondecreasing in $\alpha$ \cite{muller2013quantum}, we can deduce that $C \geq \tilde{C}$.

With all this preparation in hand, we can now state our main result.
\begin{theorem}
\label{thm:main}
Consider any $(t,n)$ QSS scheme with an adversarial model. The following are equivalent:
\begin{itemize} 
	\item The adversarial model has modified strength $\tilde{C}$.
	\item The scheme has $\epsilon$-secrecy with $\epsilon = 1 - \exp(-\tilde{C}  )$.
	\item The secret is $\epsilon$-reconstructable in terms of fidelity.
\end{itemize}
\end{theorem} 

An immediate corollary of this, given the relations discussed above, is that if the adversarial model has strength at most $C$, then the secret
is $\delta$-recoverable in diamond distance with $\delta \leq \sqrt{1 - \exp(-C)}$.

{\em Proof of Theorem \ref{thm:main}}. 
From Beny-Oreshkov duality~\cite{BeO11} between channels and complementary channels, we have
\begin{align}
\max_{\mathcal R}
F_\diamond(\mathcal{R} \circ \mathcal{N} ,\mathcal M) 
=\max_{\mathcal S}
F_\diamond(\hat{\mathcal N},\mathcal S \circ \hat{\mathcal{M}}),
\end{align}
where optimizations are over all quantum channels with appropriate input and output dimensions.
Suppose that our scheme is $\epsilon$-reconstructable in fidelity. 
By applying Beny-Oreshkov duality, we get that for any $A \subset \{1,\dots,n\}$ that
\begin{align}
\epsilon &= 1-
\min_{A : |A| \ge t} \max_{\mathcal{R}_A} F_\diamond( \mathcal{R}_A \circ \mathcal{N}_A, \mathcal{I}) \notag\\
&= 1- \min_{A : |A| \ge t} \max_{\mathcal S_A }F_\diamond(\hat{\mathcal N}_A, \mathcal S_A \circ \hat{\mathcal{I}}).
\label{apply-BeO}
\end{align}
As $\hat{\mathcal{I}}$ is the trace channel, 
$\mathcal S_A \circ \hat{\mathcal{I}}$ is without loss of generality a preparation channel $\mathcal V_{A,\sigma}$ which prepares a state $\sigma$.
Since this applies for all $A$ such that $|A| \ge t$,
it follows that the QSS scheme also has $\epsilon$-secrecy.

The crucial step in our proof relates $\max_{\sigma}
F_\diamond(\hat{\mathcal N}_A, \mathcal V_{A,\sigma})$ to the entanglement-assisted capacity of $\hat{\mathcal N}_A$
using the following lemma.
\begin{lemma}
\label{lem:heroic}
For any $A \subset \{1,\dots,n\}$, 
\begin{equation}
   \max_{\sigma } 
   F_\diamond(\hat{\mathcal N}_A, \mathcal V_{A,\sigma}) 
   = \text{e}^{-\tilde{C}_{A}},
\end{equation}
where $\tilde{C}_A = C_{\sfrac 1 2}(\hat{\mathcal N}_A)$.
\end{lemma}
In essence, Lemma \ref{lem:heroic} connects the worst-case entanglement fidelity with a variant of the entanglement-assisted capacity that arises from generalized sandwiched R\'{e}nyi divergences.

The first step in proving Lemma \ref{lem:heroic} is to show that 
 \begin{align}
& 
   F_\diamond(\hat{\mathcal N}_A, \mathcal V_{A,\sigma}) 
    = 
\min_\rho
q(\rho,\sigma)^2
\label{F2-intro}
\end{align}
where
\begin{align}
    q(\rho, \sigma)= F(
(\sqrt{\rho} \otimes I)
J
(\sqrt{\rho} \otimes I)
,
\rho\otimes \sigma)^{\nicefrac12}.
\end{align}
Here 
\begin{align}
J 
= 
(\mathds1 \otimes  \hat{\mathcal N}_A) 
\sum_{i,j} |\psi_i \>|\psi_i \> \<\psi_j|\<\psi_j| 
\end{align}
is the Choi-Jamiolkowski matrix~\cite{Choi75,jiang2013channel} of the channel $\hat{\mathcal N}_A$,
and $I$ denotes an identity matrix. 
To show \eqref{F2-intro}, 
we initially write the spectral decomposition of any density matrix $\rho$ as $\rho=\sum_{i} \lambda_i |\psi_i\>\<\psi_i|$, where $|\psi_i\>$ denotes an orthonormal basis. 
Since $\lambda_i$ are non-negative, we can write 
$\sqrt{\rho}=\sum_{i} \sqrt{\lambda_i} |\psi_i\>\<\psi_i|$.
Next, the purification of $\varrho$ is 
$\ket{\psi_\varrho}=\sum_i \sqrt{\lambda_i} |\psi_i\>|\psi_i\>$. If we trace out either the first or second part of the system of the purified state $\ket{\psi_\varrho}$, we will reconstruct the state $\rho$. 
Using this notation, note that when $\hat{\mathcal N}_A$ takes as input the state $\rho$, we have
\begin{align}
\tau =&
(\mathds1 \otimes \hat{\mathcal N}_A) 
( |\psi_\varrho \> \<\psi_\varrho|  ) 
\notag\\
=&  
\sum_{i,j} 
\sqrt {\lambda_i}
\sqrt {\lambda_j}
(\mathds1 \otimes  \hat{\mathcal N}_A) 
(|\psi_i\> |\psi_i\> \<\psi_j| \<\psi_j|   )
\notag\\
=& 
(\sqrt \rho \otimes I )
(\mathds1 \otimes \hat{\mathcal N}_A) 
(\sum_{i,j} 
|\psi_i\> |\psi_i\> \<\psi_j| \<\psi_j|   )
(\sqrt \rho \otimes I )
\notag\\
=&
(\sqrt{\rho} \otimes I)
J 
(\sqrt{\rho} \otimes I).
\end{align}
Hence we can see that 
\begin{align}
 &F_\diamond(\hat{\mathcal N}_A, \mathcal V_{A,\sigma}) 
 \notag\\
 =&
  F(
(\mathds1 \otimes \hat{\mathcal N}_A)
(\ket{\psi_\varrho}\<\psi_\varrho|)
,
\rho\otimes \sigma)
\notag\\ 
=  &
    F(
(\sqrt{\rho} \otimes I )
J 
(\sqrt{\rho} \otimes I )
,
\rho\otimes \sigma)
\notag\\
=  &
   \Bigl( \tr \sqrt{
   (\sqrt\rho\otimes \sqrt \sigma)
(\sqrt{\rho} \otimes I )
J 
(\sqrt{\rho} \otimes I )
(\sqrt\rho\otimes \sqrt \sigma)}\Bigr)
^2
\notag\\
=  &
   \Bigl( \tr \sqrt{
   ( \rho\otimes \sqrt \sigma) 
J  
( \rho\otimes \sqrt \sigma)}\Bigr)
^2
.\label{fid-to-q-halfway}
\end{align}
Using the definition of the fidelity, we note that 
 \begin{align}
 q(\rho,\sigma) &=
    \tr\sqrt{ (\rho \otimes \sigma^{\nicefrac12}) J  (\rho \otimes \sigma^{\nicefrac12})}  
    \label{q-first} \\
&= \tr \sqrt{ J ^{\nicefrac12} (\rho^2 \otimes \sigma)J ^{\nicefrac12}} \label{Eq:fidelity} \\
&= \left\| J ^{\nicefrac12} (\rho \otimes \sqrt{\sigma}) \right\|_1 .\label{Eq:fidelity2}
\end{align}
 Here in the penultimate equality, we use the fact $\tr(XJX)^{\nicefrac12} = \tr(J^{\nicefrac12} X^2 J^{\nicefrac12})^{\nicefrac12}$ for positive semi-definite $X$ and $J$. 
 From \eqref{fid-to-q-halfway} and \eqref{q-first}, we can establish \eqref{F2-intro}.

The second step in the {proof of} Lemma \ref{lem:heroic} is to show that the function 
$q(\rho,\sigma)$ 
is convex in the density matrix $\rho$ and concave in the density matrix $\sigma$.
Concavity of $q(\rho,\sigma)$ in $\sigma$ is immediate from the fact that the expression in~\eqref{Eq:fidelity} $\omega \mapsto \tr \sqrt{\omega}$ is concave and the linearity of the expression under the square root in $\sigma$.
To show convexity in $\rho$ we simply note that any norm as in~\eqref{Eq:fidelity2} is convex, and the expression inside the norm is linear in $\rho$.
Since $q(\rho, \sigma)$ is convex in $\rho$ and concave in $\sigma$, we can apply the minimax theorem~\cite{do2001introduction} to interchange the maximization and minimization, 
 in the sense that
 \begin{align}
 \max_\sigma \min_\rho q(\rho,\sigma)  
=\min_\rho \max_\sigma q(\rho,\sigma) .
\label{q-minimax}
 \end{align}

 Third, we use \eqref{q-minimax} along with the identity \eqref{F2-intro} to establish the equivalence between a fidelity and R\'{e}nyi mutual information. 

Denoting the input and output registers of $\hat { \mathcal N}_A$ as $X$ and $Y$ respectively, we see that  
\begin{align}
  \tilde I(X:Y)_\tau
    &= 
    \min_{\sigma}
    \tilde D_{\sfrac 1 2} \bigl( 
    ( \mathds1 \otimes \hat{\mathcal N}_A  )
    ( \ket{\psi_\varrho}\<\psi_\varrho| \bigl\|
    \rho \otimes \sigma)
    \bigr)
    \notag\\
    &= 
   \min_{\sigma} \left(- 
    \log F\bigl( 
     (\mathds1 \otimes \hat{\mathcal N}_A )
    ( \ket{\psi_\varrho}\<\psi_\varrho| ,
    \rho \otimes \sigma)
    \bigr) \right)
    \notag\\
    &= 
    \min_{\sigma} \left(-
    \log q(\rho , \sigma)^2\right).
\end{align}
Because $-\log$ is a monotone decreasing function, we deduce that
$\tilde I(X:Y)_\tau
= - \log \left( 
 \max_\sigma  
 q(\rho, \sigma)^2 
 \right)$. Applying the definition of the generalized entanglement assisted capacity, we get
 $ \tilde C_A
= - \log \bigl( 
 \min_\rho \max_\sigma  
 q(\rho, \sigma)^2
 \bigr)$.
 Next, the minimax result \eqref{q-minimax} implies that 
 \begin{align}
 \tilde C_A 
= - \log \bigl( 
 \max_\sigma  \min_\rho 
 q(\rho, \sigma)^2
 \bigr).     \label{q-and-capacity}
 \end{align}
 Next, from \eqref{F2-intro}, we can see that 
 $\max_\sigma F_\diamond ( \hat {\mathcal N}_A, \mathcal V_{A,\sigma})
  =  \max_\sigma  \min_\rho 
 q(\rho, \sigma)^2$.
 Hence  
\begin{align}
\exp(- \tilde C_A  )
&= 
   \max_\sigma F_\diamond ( \hat {\mathcal N}_A, \mathcal V_{A,\sigma}),
\end{align}
and the proof of Lemma \ref{lem:heroic} follows.
Putting Lemma \ref{lem:heroic} and \eqref{apply-BeO} together, we complete the proof of Theorem \ref{thm:main}. $\square$

\section{Conclusion, discussion, and open questions}
We have established that the entanglement-assisted capacity of a channel connecting the quantum secret to the quantum systems of the adversary determines both the approximate reconstructability and the approximate secrecy of a threshold QSS scheme. 
In some sense, our result can be intuitively understood from the mantra
``{\em Quantum information cannot {be} learnt without disturbing it.}''
This mantra can be used to obtain interpretations of multitude of topics in quantum theory, such as approximate quantum error correction~\cite{LNCY97,BaK02,BeO10,Tys10,ouyang2014permutation}, monogamy of entanglement~\cite{tomamichel2013monogamy}, and the quantum information of black hole evaporation~\cite{hawking-RevModPhys.93.035002}.
Particularly for quantum error correction, the encoding map in a QSS scheme takes the quantum secret to a quantum error correction code, and the approximate reconstructability of the secret is precisely the approximate reconstructability of the code. 
In this regard, our theorem implies that, 
if the adversaries trying to learn the secret have access to a channel with entanglement-assisted capacity of $C$, then there exists a decoding operation that reconstructs the secret up to an error of $\delta$, quantified in terms of the diamond distance, where $\delta \le \sqrt{1-\exp(-C  )}$.
It remains an open question as to how  different types of capacities {other} than the entanglement-assisted capacity influences the theory of approximate QSS.

\section*{Acknowledgements}
YO and MT are supported by the Quantum Engineering Programme grant NRF2021-QEP2-01-P06, and the National Research Foundation, Prime Minister’s Office, Singapore and the Ministry of Education, Singapore under the Research Centres of Excellence program. YO also acknowledges support from EPSRC (Grant No. EP/W028115/1). This research was supported by the Australian Research Council (ARC) Discovery Project (DP200102273) and ARC Centre of Excellence for Engineered Quantum Systems (EQUS,CE170100009). JR is supported by a Westpac Bicentennial Foundation Research Fellowship.
BCS acknowledges funding from the Natural Sciences and Engineering Research Council of Canada.

\appendix
\section{Supplemental Material}

 First we define some notation. Given a Hilbert space $\mathscr H$, let $|\mathscr H|$ denote its dimension.
We restrict our attention to finite dimensional Hilbert spaces.
Let $\textsf{M}(\mathscr H)$ denote the set of matrix representations of linear operators on Hilbert space $\mathscr H$. 
 Let $\textsf{D}(\mathscr H)$ denote the set of operators in $\textsf{M}(\mathscr H)$ that have unit trace and are positive semidefinite.
 A quantum channel is a completely positive and trace preserving map from 
 $\textsf{M}(\mathscr H)$ to $\textsf{M}(\mathscr K)$ where $\mathscr H$ and $\mathscr K$ are Hilbert spaces.
 We use the shorthand ($\mathcal N$ \textsf{CPT}) to indicate that $\mathcal N$ is a quantum channel.

\begin{proof}[Proof of (6) in the main manuscript]
Note that for a channel $\mathcal F: \textsf{M}(\mathscr H ) \to \textsf{M}(\mathscr H)$,  
\begin{align}
    F ( \mathcal F, \mathds1 ) 
   =& \min_{\substack{|\psi\> \in \mathscr H \otimes \mathscr H \\
    \| |\psi\> \|= 1\\
    }}
    F( |\psi\>\<\psi|, (\mathcal I \otimes \mathcal F) ( |\psi\>\<\psi| ) ).
\end{align}
Now for any pure state $|\psi\>\<\psi|$ and mixed state $\sigma$, we have
\begin{align}
    F(|\psi\>\<\psi|, \sigma)
    = \<\psi| \sigma |\psi\>
\end{align}
From the Fuchs-van de Graaf inequalities we have
\begin{align}
\bigl(1- \frac{1}{2}\| |\psi\>\<\psi| - \sigma \|_1\bigr)^2
&\le F(|\psi\>\<\psi|, \sigma)
\notag\\
F(|\psi\>\<\psi|, \sigma) &\le 1- \frac{1}{4}\| |\psi\>\<\psi| - \sigma \|_1^2.
\end{align}
We thereby deduce that 
\begin{align}
&F ( \mathcal F, \mathds1 ) \notag\\
\le& 1 - \max_{\substack{|\psi\> \in \mathscr H \otimes \mathscr H \notag\\
    \| |\psi\> \| = 1\\
    }}
\frac{1}{4}
\left\| |\psi\>\<\psi| - 
(\mathcal I \otimes \mathcal F)(|\psi\>\<\psi|) 
\right\|_1^2
\notag\\
=&  1- \frac{1}{4}
\left\| \mathds1 -\mathcal F \right\|_\diamond^2 \notag\\
=&  1- D_\diamond( \mathds1, \mathcal F)^2,
\end{align}
and
\begin{align}
&F ( \mathcal F, \mathds1 ) \notag\\
\ge& (1 - \max_{\substack{|\psi\> \in \mathscr H \otimes \mathscr H \notag\\
    \| |\psi\> \| = 1\\
    }}
\frac{1}{2}\| |\psi\>\<\psi| - 
(\mathcal I \otimes \mathcal F)(|\psi\>\<\psi|) 
\|_1)^2
\notag\\
=&  (1- \frac{1}{2}\| \mathds1 -\mathcal F\|_\diamond)^2 \notag\\
=& ( 1- D_\diamond( \mathds1, \mathcal F))^2.
\end{align}
Hence,
\begin{align}
 ( 1- D_\diamond( \mathds1, \mathcal F))^2
 \le F ( \mathcal F, \mathds1 ) \le
  1- D_\diamond( \mathds1, \mathcal F)^2.
\end{align}
For a tighter lower bound, note that \cite[Lemma 9.1.1]{Wilde}
\begin{align}
\frac{1}{2} \| \psi\>\<\psi| - \sigma \|_1 
=
\max_{0\le P\le I} \operatorname{tr}P(|\psi\>\<\psi|-\sigma),
\end{align}
and by picking $P=|\psi\>\<\psi|$, we get
\begin{align}
\frac{1}{2} \| \psi\>\<\psi| - \sigma \|_1 
\ge 
1- \<\psi|\sigma|\psi\> = 1-F(|\psi\>\<\psi|,\sigma),
\end{align}
and hence 
\begin{align}
 1- D_\diamond( \mathds1, \mathcal F)
 \le F ( \mathcal F, \mathds1 ) \le
  1- D_\diamond( \mathds1, \mathcal F)^2,
\end{align}
and this proves (6) in the main manuscript. 
\end{proof}


\bibliographystyle{ieeetr}
\bibliography{qsecret}

\begin{thebibliography}{10}

\bibitem{BB84}
C.~H. Bennett and G.~Brassard, ``{Quantum cryptography: Public key distribution
  and coin tossing},'' in {\em Proc. IEEE International Conference on
  Computers, Systems and Signal Processing}, vol.~175, New York, 1984.

\bibitem{Eke91}
A.~K. Ekert, ``Quantum cryptography based on {B}ell's theorem,'' {\em Phys.
  Rev. Lett.}, vol.~67, pp.~661--663, Aug 1991.

\bibitem{hillery_quantum_1999}
M.~Hillery, V.~Buzek, and A.~Berthiaume, ``Quantum secret sharing,'' {\em Phys.
  Rev. A}, vol.~59, pp.~1829--1834, Mar. 1999.
\newblock arXiv: quant-ph/9806063.

\bibitem{CGL99}
R.~Cleve, D.~Gottesman, and H.-K. Lo, ``How to share a quantum secret,'' {\em
  Phys. Rev. Lett.}, vol.~83, pp.~648--651, Jul 1999.

\bibitem{KnL97}
E.~Knill and R.~Laflamme, ``{Theory of quantum error-correcting codes},'' {\em
  Phys. Rev. A}, vol.~55, pp.~900--911, Feb. 1997.

\bibitem{markham2008}
D.~Markham and B.~C. Sanders, ``Graph states for quantum secret sharing,'' {\em
  Phys. Rev. A}, vol.~78, p.~042309, Oct 2008.

\bibitem{KFMS10}
A.~Keet, B.~Fortescue, D.~Markham, and B.~C. Sanders, ``Quantum secret sharing
  with qudit graph states,'' {\em Phys. Rev. A}, vol.~82, p.~062315, Dec 2010.

\bibitem{quantum-byzantine-prl}
M.~Fitzi, N.~Gisin, and U.~Maurer, ``Quantum solution to the byzantine
  agreement problem,'' {\em Phys. Rev. Lett.}, vol.~87, p.~217901, Nov 2001.

\bibitem{Ouyang_QSS_2017}
Y.~Ouyang, S.-H. Tan, L.~Zhao, and J.~F. Fitzsimons, ``Computing on quantum
  shared secrets,'' {\em Physical Review A}, vol.~96, no.~5, p.~052333, 2017.

\bibitem{Got00}
D.~Gottesman, ``Theory of quantum secret sharing,'' {\em Phys. Rev. A},
  vol.~61, p.~042311, Mar 2000.

\bibitem{imai_quantum_2003}
H.~Imai, J.~Mueller-Quade, A.~C.~A. Nascimento, P.~Tuyls, and A.~Winter, ``A
  quantum information theoretical model for quantum secret sharing schemes,''
  {\em Quantum Inf. Comput.}, vol.~5, pp.~69--80, 2005.
\newblock arXiv: quant-ph/0311136.

\bibitem{HBB99}
M.~Hillery, V.~Bu\ifmmode~\check{z}\else \v{z}\fi{}ek, and A.~Berthiaume,
  ``Quantum secret sharing,'' {\em Phys. Rev. A}, vol.~59, pp.~1829--1834, Mar
  1999.

\bibitem{nikova2006threshold}
S.~Nikova, C.~Rechberger, and V.~Rijmen, ``Threshold implementations against
  side-channel attacks and glitches,'' in {\em International conference on
  information and communications security}, pp.~529--545, Springer, 2006.

\bibitem{Par70}
J.~L. Park, ``The concept of transition in quantum mechanics,'' {\em Found.
  Phys.}, vol.~1, pp.~23--33, Mar 1970.

\bibitem{WZ82}
W.~K. Wootters and W.~H. Zurek, ``A single quantum cannot be cloned,'' {\em
  Nature}, vol.~299, no.~5886, pp.~802--803, 1982.

\bibitem{crepeau2005approximate}
C.~Cr{\'e}peau, D.~Gottesman, and A.~Smith, ``Approximate quantum
  error-correcting codes and secret sharing schemes,'' in {\em Annual
  International Conference on the Theory and Applications of Cryptographic
  Techniques}, pp.~285--301, Springer, 2005.

\bibitem{imai2005information}
H.~Imai, J.~M{\"u}ller-Quade, A.~C. Nascimento, P.~Tuyls, and A.~Winter, ``An
  information theoretical model for quantum secret sharing,'' {\em Quant. Inf.
  Comput.}, vol.~5, no.~1, pp.~69--80, 2005.

\bibitem{Ogawa_yamamoto_secret}
T.~Ogawa, A.~Sasaki, M.~Iwamoto, and H.~Yamamoto, ``Quantum secret sharing
  schemes and reversibility of quantum operations,'' {\em Phys. Rev. A},
  vol.~72, p.~032318, Sep 2005.

\bibitem{Spekkens_complementarity}
D.~Kretschmann, D.~W. Kribs, and R.~W. Spekkens, ``Complementarity of private
  and correctable subsystems in quantum cryptography and error correction,''
  {\em Phys. Rev. A}, vol.~78, p.~032330, Sep 2008.

\bibitem{hayden_approximate_2020}
P.~Hayden and G.~Penington, ``Approximate {Quantum} {Error} {Correction}
  {Revisited}: {Introducing} the {Alpha}-{Bit},'' {\em Commun. Math. Phys.},
  vol.~374, pp.~369--432, Mar. 2020.

\bibitem{DeS03}
I.~Devetak and P.~W. Shor, ``{The Capacity of a Quantum Channel for
  Simultaneous Transmission of Classical and Quantum Information},'' {\em
  Commun. Math. Phys.}, vol.~256, no.~2, pp.~287--303, 2005.

\bibitem{muller2013quantum}
M.~M{\"u}ller-Lennert, F.~Dupuis, O.~Szehr, S.~Fehr, and M.~Tomamichel, ``On
  quantum {R}{\'e}nyi entropies: A new generalization and some properties,''
  {\em J. Math. Phys.}, vol.~54, no.~12, p.~122203, 2013.

\bibitem{wilde2014strong}
M.~M. Wilde, A.~Winter, and D.~Yang, ``Strong converse for the classical
  capacity of entanglement-breaking and hadamard channels via a sandwiched
  r{\'e}nyi relative entropy,'' {\em Commun. Math. Physics}, vol.~331, no.~2,
  pp.~593--622, 2014.

\bibitem{gupta2015multiplicativity}
M.~K. Gupta and M.~M. Wilde, ``Multiplicativity of completely bounded p-norms
  implies a strong converse for entanglement-assisted capacity,'' {\em Commun.
  Math. Phys.}, vol.~334, no.~2, pp.~867--887, 2015.

\bibitem{BeO11}
C.~B\'{e}ny and O.~Oreshkov, ``{Approximate simulation of quantum channels},''
  {\em Phys. Rev. A}, vol.~84, p.~022333, Aug. 2011.

\bibitem{Choi75}
M.-D. Choi, ``{Completely positive linear maps on complex matrices},'' {\em
  Linear Algebra and its Applications}, vol.~10, no.~3, pp.~285--290, 1975.

\bibitem{jiang2013channel}
M.~Jiang, S.~Luo, and S.~Fu, ``Channel-state duality,'' {\em Phys. Rev. A},
  vol.~87, no.~2, p.~022310, 2013.

\bibitem{do2001introduction}
M.~do~Ros{\'a}rio~Grossinho and S.~A. Tersian, {\em An introduction to minimax
  theorems and their applications to differential equations}, vol.~52.
\newblock Springer Science \& Business Media, 2001.

\bibitem{LNCY97}
D.~W. Leung, M.~A. Nielsen, I.~L. Chuang, and Y.~Yamamoto, ``{Approximate
  quantum error correction can lead to better codes},'' {\em Phys. Rev. A},
  vol.~56, p.~2567, 1997.

\bibitem{BaK02}
H.~Barnum and E.~Knill, ``{Reversing quantum dynamics with near-optimal quantum
  and classical fidelity},'' {\em J. Math. Phys.}, vol.~43, p.~2097, Jan. 2002.

\bibitem{BeO10}
C.~B\'{e}ny and O.~Oreshkov, ``{General Conditions for Approximate Quantum
  Error Correction and Near-Optimal Recovery Channels},'' {\em Phys. Rev.
  Lett.}, vol.~104, p.~120501, Mar. 2010.

\bibitem{Tys10}
J.~Tyson, ``{Two-sided bounds on minimum-error quantum measurement, on the
  reversibility of quantum dynamics, and on maximum overlap using directional
  iterates},'' {\em J. Math. Phys.}, vol.~51, p.~92204, June 2010.

\bibitem{ouyang2014permutation}
Y.~Ouyang, ``Permutation-invariant quantum codes,'' {\em Phys. Rev. A},
  vol.~90, p.~062317, Dec 2014.

\bibitem{tomamichel2013monogamy}
M.~Tomamichel, S.~Fehr, J.~Kaniewski, and S.~Wehner, ``A
  monogamy-of-entanglement game with applications to device-independent quantum
  cryptography,'' {\em New J. Phys.}, vol.~15, no.~10, p.~103002, 2013.

\bibitem{hawking-RevModPhys.93.035002}
A.~Almheiri, T.~Hartman, J.~Maldacena, E.~Shaghoulian, and A.~Tajdini, ``The
  entropy of hawking radiation,'' {\em Rev. Mod. Phys.}, vol.~93, p.~035002,
  Jul 2021.

\bibitem{Wilde}
M.~M. Wilde, {\em {From Classical to Quantum Shannon Theory}}.
\newblock Cambridge University Press, 2013.

\end{thebibliography}

\end{document}